\documentclass[10pt,prd,tightenlines,twocolumn,superscriptaddress]{revtex4}

\usepackage{amsmath,amssymb}

\usepackage{color}
\usepackage{graphicx}
\definecolor{darkblue}{rgb}{0,0,0.7}
\definecolor{darkred}{rgb}{0.7,0,0}
\usepackage[unicode, colorlinks, citecolor=darkblue, linkcolor=darkred, urlcolor=blue]{hyperref}

\begin{document}

\title{Sensitivity of laser gravitational-wave detectors with stable double-pumped optical spring }

\author{Andrey A. Rakhubovsky \footnote{Electronic address: rkh@hbar.phys.msu.ru}}
\affiliation{Physics Department, Moscow State University, Moscow 119992 Russia}

\author{Sergey P. Vyatchanin}
\affiliation{Physics Department, Moscow State University, Moscow 119992 Russia}
\date{\today}

\begin{abstract}

	We analyze the sensitivity of gravitational-wave antenna with stable double optical spring created by two independent pumps. We investigate regime of three close eigen frequencies (roots of characteristic equation) which appears to provide more wide frequency band in which sensitivity of antenna can beat Standard Quantum Limit (SQL) than previously considered regime with two close eigen frequencies. We take into account optical losses and show that they do not degrade sensitivity significantly. We also demonstrate possible application of considered regime to Einstein Telescope. 

\end{abstract}

\maketitle

\section{Introduction}

Currently the search for gravitational radiation from astrophysical
sources is carried out with the first-generation Earth-based laser
interferometers (LIGO in USA \cite{1992_LIGO,2006_LIGO_status,website_LIGO}, VIRGO in Italy
\cite{2006_VIRGO_status,website_VIRGO}, GEO-600 in Germany
\cite{2006_GEO-600_status,website_GEO-600}, TAMA-300 in Japan
\cite{2005_TAMA-300_status,website_TAMA-300} and ACIGA in Australia
\cite{2006_ACIGA_status,website_ACIGA}). The development of the
second-generation GW detectors (Advanced LIGO
\cite{2002_Adv_LIGO_config,website_Adv_LIGO}, Advanced Virgo
\cite{website_Advirgo}, GEO-HF \cite{GEO-HF} and LCGT
\cite{2006_LCGT_status}) is well underway.

The sensitivity of the first-generation detectors is limited by noises sources of various nature: seismic and suspension thermal noise at low frequencies (below $\sim 50$ Hz), thermal noise in suspensions, bulks and coatings of the mirrors  ($\sim 50-200$ Hz), photon shot noise (above $\sim 200$ Hz). It is expected that the sensitivity of the second-generation detectors will be ultimately limited by the noise of quantum nature arising due to Heisenberg's uncertainty principle \cite{1968_SQL,1975_SQL,1977_SQL,1992_quant_meas} over most of the frequency range of interest. The optimum between  measurement noise (photon shot noise) and back-action noise (radiation pressure noise) is called the Standard Quantum Limit (SQL) This level is expected to be reached in the forthcoming second generation of large-scale laser-interferometric gravitational-wave detectors. Third generation detectors, such as the Einstein Telescope \cite{et_punturo2010} aim to significantly surpass the SQL over a wide frequency range \cite{Hild10}.

The most promising methods to overcome the SQL rely on the implementation of optical (ponderomotive) rigidity \cite{64a1BrMiVMU,67a1BrMaJETP,78BrBook, 1992_quant_meas} which effectively turns the test masses of a gravitational-wave detector into harmonic oscillators producing gain in sensitivity \cite{99a1BrKhPLA,01a1KhPLA,01ChenPRD,02ChenPRD,05a1LaVyPLA,06a1KhLaVyPRD}. A single optical spring always causes instability. The instability (negative damping for mechanical degree of freedom) can be compensated by incorporating a linear feedback control loop and in the ideal case (no additional noise is introduced by the feedback) it would not modify the noise spectrum of a GW detector \cite{02ChenPRD}.  In practice, however, the need for control gain at frequencies inside the detection band can cause undesirable complexity in the control system or can introduce additional classical noise.

\begin{figure}[t]
	\includegraphics[width=\linewidth]{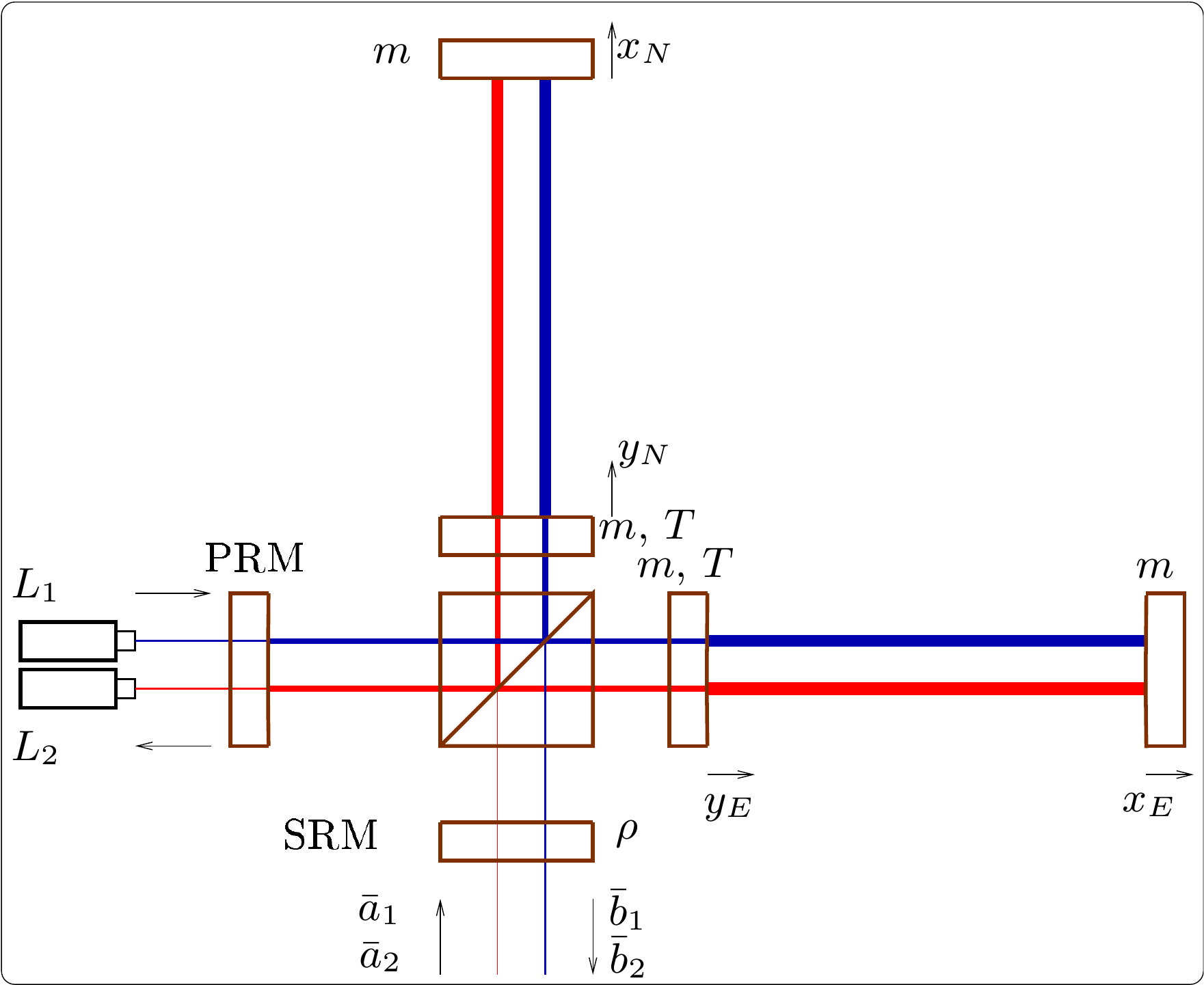}
	\caption{Scheme of an Advanced LIGO interferometer pumped by two lasers. The main laser is detuned to give a positive optical rigidity and negative damping (it is tuned on the right slope of resonance curve)  while the auxiliary  laser is detuned to give negative optical rigidity and positive  damping.
	}
	\label{fig:AdvLIGO}
\end{figure}

An alternative way to suppress the instability was proposed \cite{08ChenPRD} and experimentally demonstrated \cite{07CorbitPRL}, by injecting a second carrier field from the bright port (see Fig.~\ref{fig:AdvLIGO}) in order  to create a relatively small additional negative rigidity component, thus leaving the total rigidity (of both lasers together) positive, but at the same time to introduce a relatively large additional positive damping component to make total damping positive. The main purpose of the second carrier is to create a second optical spring that forms a stable optical spring together with the first one --- even though each individual optical spring, acting alone, would be unstable. Both carriers are assumed to have different  polarizations (or carriers' frequencies to differ from each other by large enough value), so that there is no direct coupling between the two pump field (although they both directly couple to the mirrors). 

A simple criterion for the stability of an optical spring and its application to the double resonance regime was presented in \cite{11Rakhubovsky}. It allows to use attractive regimes of double resonance \cite{05a1LaVyPLA,06a1KhLaVyPRD,11Rakhubovsky} or negative inertia \cite{10KhLIGO} in stable variants with no need to use feedback loops.

In this paper we further analyze the sensitivity of Advanced LIGO (aLIGO) interferometer with a double pump using approach \cite{08ChenPRD}.  To obtain more wide sensitivity curve we use the regime of {\em three} close roots of characteristic equation instead of double resonance regime. However this regime demands smaller interferometer relaxation rates, close to limit caused by optical losses, so there is need to consider optical losses too. Analysis shows that presence of losses doesn't affect the sensitivity curves much. Finally we apply this regime to the parameters of Einstein Telescope \cite{Hild10} and show that use of stable double optical spring makes possible to make an improvement to sensitivity in small frequency range.

\section{Output analysis}\label{dpumpOut}

We consider the balanced interferometer similar to aLIGO configuration (see Fig.~\ref{fig:AdvLIGO}) pumped by two lasers. We assume that vacuum fluctuations input through dark port (no squeezing), Fabry-Perot cavities in arms are identical and have no optical losses. The masses of input and end mirrors in arms are equal to $m$. The mean frequencies of each pump  $\omega_{1,2}$ are equal to one of the eigen frequencies of each  cavity. Here and below subscripts $_{1,2}$ refer to main and auxiliary pumps. We also introduce the notations:
\begin{subequations}
\begin{align}
\label{gamma0}
 e^{i\Omega L/c}& \simeq 1+\frac{i\Omega L}{c},
 	\quad \gamma_0=\frac{cT^2}{4L},\\
\phi_{1,2} &= \frac{(\omega_{1,2}+\Omega) \ell}{c}\simeq\frac{\omega_{1,2} \ell}{c}, \\
\label{Gamma}
\Gamma_{1,2} &= \gamma_0\,\frac{1-\rho^2}{1+2\rho\cos2\phi_{1,2}+\rho^2}\,,\\
\label{Delta}
\Delta_{1,2} & = \gamma_0\,\frac{2\rho\sin 2\phi_{1,2}}{1+2\rho\cos2\phi_{1,2}+\rho^2}\,,\
\end{align}
\end{subequations}
Here $L$ is the distance between the mirrors in arms  ($4$~km for aLIGO), $\ell$ is the distance (several meters) between SR mirror and the input mirrors in Fabry-Perot cavities, due to strong inequality $l\ll L$ we assume that $\phi_{1,2}$ do not depend on $\Omega$. We assume that arm cavities are tuned in resonance, $\gamma_0$ is the relaxation rate of single Fabry-Perot cavity in arm, $T$ is amplitude transmittance of input mirrors in arms, $\rho$ is amplitude reflectivity of SR mirror, $\Delta_{1,2}$ are the detunings introduced by displacement shift of SR mirror, $\Gamma_{1,2}$ is the relaxation rates of the  differential mode of the interferometer for each pump. It is worth underlying  that detunings $\Delta_{1,2}$ and relaxation rates $\Gamma_{1,2}$ may be different for each pump. 

We start from equations for output quadratures amplitudes $b_{1,2}^{(1,2)}$  expressed in terms of input quadratures amplitudes $a_{1,2}^{(1,2)}$ in frequency domain (superscripts $^{(1,2)}$ refer to different quadratures, see details and notations in Appendix~\ref{notations})
\begin{subequations}
\label{starteq}
\begin{align}
  b_1^{(1)} &=  a_1^{(1)} +U_1 \cdot \mu\Omega^2\Psi \cdot \mathcal{F},\\
  b_1^{(2)} &=  a_1^{(2)} +U_2 \cdot \mu\Omega^2\Psi \cdot \mathcal{F},\\
  b_2^{(1)} &=  a_2^{(1)} +U_3 \cdot \mu\Omega^2\Psi \cdot \mathcal{F},\\
  b_2^{(2)} &=  a_2^{(2)} +U_4 \cdot \mu\Omega^2\Psi \cdot \mathcal{F},\\
  \label{F}
 \mathcal{F}  &\equiv  
	 	  T_1 a_{1}^{(1)} +T_2	 a_{1}^{(2)}+ T_3 a_{2}^{(1)} + T_4	 a_{2}^{(2)}
	 	+ \frac{\sqrt2\,h}{h_\text{SQL}}\, ,\\
 \Psi^{-1} &\equiv -\mu\Omega^2+K_{1}+K_{2},\\ 
	K_{1,2} &\equiv \frac{2\Delta_{1,2}I_{1,2}\omega_{1,2}}{cL
	\big[ (\Gamma_{1,2}-i\Omega)^2+\Delta_{1,2}^2\big]}\,,\\
 h_\text{SQL}&\equiv \sqrt\frac{2\hslash}{\mu\Omega^2\,L^2},\quad \mu\equiv \frac{m}{4}\,.
\end{align}
\end{subequations}
Here $\Psi$ is susceptibility of mechanical degree of freedom accounting optical rigidities $K_{1,2}$ introduced by pumps 1 and 2, $I_{1,2}$ are mean powers circulating in arms (pumped by main and auxiliary lasers), $\mu$ is reduced mass, $\Omega$ is spectral frequency, $h$ is dimensionless gravitational metric perturbation normalized by SQL perturbation  $h_\text{SQL}$. The term $\mathcal{F} $ in formulas (\ref{starteq}) describes fluctuational (back action) and signal forces, while the first terms in right parts describe measurement errors. The coefficients $U_i,\ T_i$  are the following:
\begin{subequations}
\label{UT}
\begin{align}
U_1  &\equiv -\sqrt{\mathcal{Q}_1}\Theta_1\,\Delta_1,\quad 
	U_2 \equiv \sqrt{\mathcal{Q}_1}\Theta_1\left[\Gamma_1-i\Omega\right], \\
U_3  &\equiv -\sqrt{\mathcal{Q}_2}\Theta_2\,\Delta_2,\quad 
	U_4 \equiv \sqrt{\mathcal{Q}_2}\Theta_2\left[\Gamma_2-i\Omega\right],	\\
T_1 &\equiv \sqrt{\mathcal{Q}_1}\Theta_1^*
		\left[\Gamma_1+i\Omega\right],\quad
	T_2 \equiv \sqrt{\mathcal{Q}_1}\Theta_1^*\,\Delta_1,\\
T_3 &\equiv \sqrt{\mathcal{Q}_2}\Theta_2^*
		\left[\Gamma_2+i\Omega\right],\quad
	T_4 \equiv \sqrt{\mathcal{Q}_2}\Theta_2^*\,\Delta_2\,,\\
\mathcal{Q}_{1,2}&  \equiv   \frac{2\Gamma_{1,2}|K_{1,2}|}{\Delta_{1,2}\,\mu\Omega^2}\times
	\frac{1}{\big| (\Gamma_{1,2}-i\Omega)^2+\Delta_{1,2}^2\big|},\\
	 \Theta_{1,2} &\equiv \sqrt
	 \frac{(\Gamma_{1,2}+i\Omega)^2+\Delta_{1,2}^2}{(\Gamma_{1,2}-i\Omega)^2+\Delta_{1,2}^2}\,.	
\end{align}
\end{subequations}
The equations (\ref{starteq},\ref{UT}) slightly differ from equations used in \cite{08ChenPRD} due to different definition of quadrature amplitudes (see (\ref{b12}, \ref{bara})).

In experiment  homodyne detector of output field measures arbitrary combination of quadratures in each channel defined by homodyne angles $\zeta_{1,2}$:
\begin{subequations}
\label{b12psi}
\begin{align}
 j_1 & = b_1^{(1)} \cos\zeta_1 +b_1^{(2)} \sin\zeta_1,\\
 j_2 & = b_2^{(1)} \cos\zeta_2 +b_2^{(2)} \sin\zeta_2.
\end{align}
\end{subequations}

For input field in vacuum state quadratures $a_i^{(j)}$ do not correlate to each other and their single-sided spectral densities  \cite{02a1KiLeMaThVyPRD} are equal to
\begin{equation}
\label{Sd}
 S_{a_i^{(j)}}(\Omega) =1
\end{equation}

In order to find condition under which sensitivity is better than SQL one may apply the following semiqualitative consideration. 

From equations (\ref{starteq}) we see that if susceptibility is large enough ($|\mu\Omega^2\Psi|\gg 1$) the back action force prevails over measurement errors and mainly defines sensitivity and spectral density $S_\text{hBA}$ of noise recalculated to dimensionless metric $h$ is (see Eq.~\ref{F}):
\begin{align}
	\notag
 \xi_\text{BA}^2&\equiv \frac{S_\text{hBA}}{h_\text{SQL}^2} =
 \frac{|T_1|^2+|T_2|^2+|T_3|^2+|T_4|^2}{2}=\\
 \label{ShK1}
 &=\frac{\Gamma_1\,|K_1|}{\Delta_1\,\mu\Omega^2} \left[
 	\frac{|\Gamma_1+i\Omega|^2 +\Delta_1^2}{\big| (\Gamma_{1}-i\Omega)^2+\Delta_{1}^2\big|}
 	\right]+\\
 &\qquad +\frac{\Gamma_2\,|K_2|}{\Delta_2\,\mu\Omega^2} \left[
 	\frac{|\Gamma_2+i\Omega|^2 +\Delta_2^2}{\big| (\Gamma_{2}-i\Omega)^2+\Delta_{2}^2\big|}
 	\right]\,.
	\notag
\end{align}

It is obvious to assume that close to resonance (when $|\mu\Omega^2\Psi|\gg 1$) we have approximation: $|K_1+K_2|/\mu\Omega^2\simeq 1$. Hence, from (\ref{ShK1})  one may conclude that this term will be less than unity (i.e. sensitivity is better than SQL, $S_\text{hBA}\ll h_\text{SQL}^2$) in case of small optical relaxation rates, i.e. 
\begin{align}
\label{cond1}
\Gamma_1\ll \Delta_1 ,\quad \Gamma_2\ll \Delta_2
\end{align}
If rigidity $|K_1|$ introduced by main pump is  larger than $|K_2|$ we should keep in mind that first inequality ($\Gamma_1\ll \Delta_1$) in (\ref{cond1}) is more important than second one ($\Gamma_2\ll \Delta_2$).

\begin{table}[ht]
\caption{Planned parameters of Advanced LIGO}\label{table:LIGO}
\begin{tabular}{c|c}
Parameter & Value \\
\hline
Arm length, $L$ & $4$~km\\
Mass of each mirror, $m$ &$40$~kg\\
ITM amplitude transmittance, $T$ & $\sqrt {0.005}\ \big(\sqrt{0.015}\big)$\\
Relaxation rate of single FP cavity, $\gamma_0$ (\ref{gamma0})& $94$~s$^{-1}$\\
SRM amplitude reflectivity $\rho$ & $\sqrt{0.95}\ \big(\sqrt{0.8}\big)$\\
Optical wavelength, $\lambda$ & $1064$~nm\\
Optical losses $A^2$ in each arm per roundtrip & $ 10^{-5}$\\
\hline
\end{tabular}
\end{table}

For frequencies far from resonance susceptibility decreases and when $|\mu\Omega^2\Psi|\simeq \Delta_1/\Gamma_1$ back action and measurement noises will be approximately equal to each other. Hence, one may get better than SQL sensitivity in bandwidth where susceptibility is high enough. 
 
In more general case formula for sensitivity is more complicated than \eqref{ShK1}. If we measure, for example, only quadrature $b_1^{(1)}$ the sensitivity is equal to
\begin{align}
 \xi_1^{\zeta} &=\frac{S_h^{(\zeta)}}{h_{SQL}^2} =\frac{1}{2}
 	\left|\frac{\cos\zeta}{\mu\Omega^2\Psi\big(U_1\cos\zeta+U_2\sin\zeta\big)}+T_1\right|^2 +
 	\nonumber\\
 	\label{xi1zeta}
 	&\quad +\frac{1}{2}
 	\left|\frac{\sin\zeta}{\mu\Omega^2\Psi\big(U_1\cos\zeta+U_2\sin\zeta\big)}+T_2\right|^2 +\\
 	\nonumber
 	&\qquad + \frac{|T_3|^2}{2} + \frac{T_4|^2}{2}.
\end{align}
Note that for single-pumped interferometer regime with small relaxation rate is not the attractive one because it may provide SQL overcoming only in small bandwidth (decreasing with decrease of $\Gamma_1$). That is why in aLIGO relatively small transmittance $\sqrt{1-\rho^2}$ is planned. From formulas (\ref{Gamma}, \ref{Delta}) and parameters listed in Table~\ref{table:LIGO} one may easy estimate that ratio $\Gamma_{1}/\Delta_1$ is restricted by $\Gamma_1/\Delta_1 \ge 0.025$. Below we do not take into account this restriction assuming that parameters $\Gamma_{1,2},\ \Delta_{1,2}$ may be chosen arbitrary and transmittances $T$ and $\sqrt{1-\rho^2}$ may be changed. 

\begin{figure}[t]

\includegraphics[width = \linewidth]{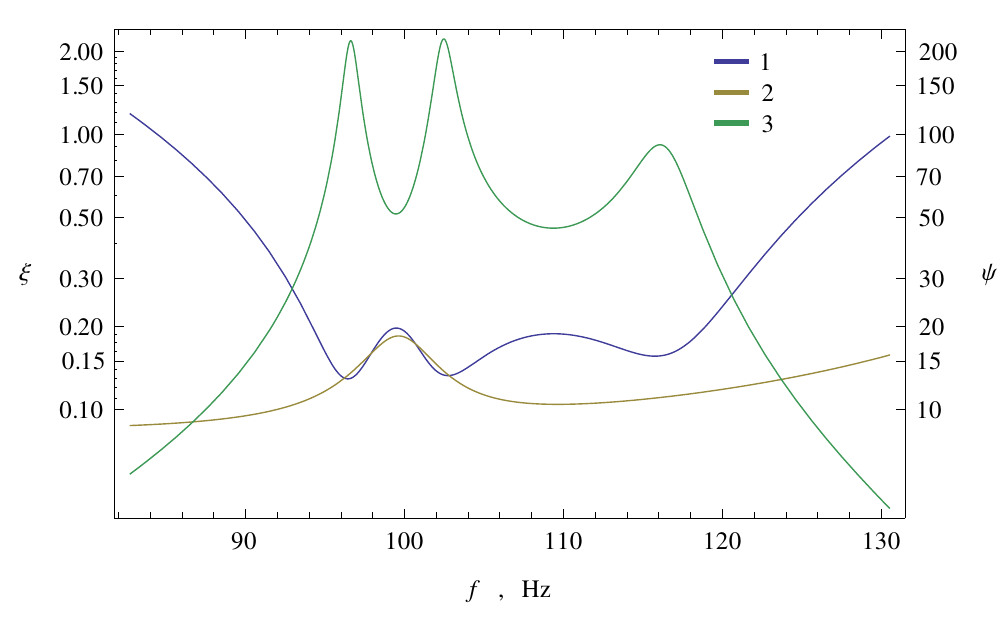}
\caption{Trace 1: sensitivity $\xi^{\zeta=0}(f)$ \eqref{xi1zeta}; 2: sensitivity $\xi_\text{BA}(f)$ (\ref{ShK1}) defined only by back action; 3: corresponding dimensionless susceptibility $\psi = \mu (2 \pi f)^2 \Psi$ for parameters~\eqref{param1}. }

\label{fig:plot1}
\end{figure}

\subsection{Close resonances regime}

In double-pumped interferometer one may manipulate by value of susceptibility $\Psi$ in wide range due to sophisticated frequency dependence of optical string. It is important that this could be done avoiding {\em instability} (it means that imaginary parts of roots of characteristic equation $\Psi^{-1}=0$ should have {\em negative} signs) \cite{11Rakhubovsky}. One of possibilities is the usage of so called double resonance regime \cite{01a1KhPLA, 05a1LaVyPLA, 06a1KhLaVyPRD} which allows to obtain two close or coinciding resonance frequencies. 

In order to obtain more wide bandwidth of high susceptibility one may use the case of  {\em three} resonance frequencies being relatively close to each other. As example we present the plots of sensitivity and susceptibility on Fig.~\ref{fig:plot1} for such particular case. For these plots we used the following parameters
\begin{subequations}
\label{param1}
 \begin{align}
  I_1 &\simeq 731\, \text{kW},\quad 
  \Gamma_1 \simeq  3.06\, \text{s}^{-1},\quad \Delta_1\simeq 961\, \text{s}^{-1},\\
  I_2 & \simeq 1.06\, \text{kW},\quad 
  	\Gamma_2 \simeq 15.3\, \text{s}^{-1},\quad \Delta_2 \simeq  -626\, \text{s}^{-1}.
 \end{align}
\end{subequations}
For these parameters the complex roots of characteristic equation  are equal to 
\begin{subequations}
\label{roots1}
 \begin{align} 
       \Omega_1 &\simeq \pm 2\pi\times 96.6 - i\,0.49\, \text{s}^{-1},\\
       \Omega_2 &\simeq \pm 2\pi\times 102.4 - i\,0.63\, \text{s}^{-1},\\
       \Omega_3 &\simeq \pm 2\pi\times 116.3 - i\,1.81\, \text{s}^{-1}.
 \end{align}
\end{subequations}
The real part of roots are eigen frequencies and imaginary parts describe relaxation (negative sign of imaginary parts corresponds to relaxation, positive one --- to instability). All roots (\ref{roots1}) are stable. We used procedure of root calculation \cite{11Rakhubovsky} allowing to have {\em stable} roots (with negative imaginary parts) avoiding instability.

Strictly speaking we should use both outputs measuring the quadratures (\ref{b12psi}) and taking their weighted sum (the corresponding procedure is given in Appendix to \cite{08ChenPRD}). However, for our case (\ref{param1}) the circulating powers $I_1$ and $I_2$ differ to each other so dramatically that useful information in port $2$ is negligible small and it is quite enough to measure only quadrature $b^\zeta_1$. In particular, on Fig.~\ref{fig:plot1} we present sensitivity (\ref{xi1zeta}) for the case $\zeta=0$. 

The presented example demonstrates the main advantage of close resonances regime: high sensitivity (about $10$ times better than SQL) in bandwidth as wide as about one half of mean frequency. 

Note that sensitivity curve presented on Fig.~\ref{fig:plot1} may be easy tuned to another frequency range by control powers $I_{1,2}$, detunings $\Delta_{1,2}$ and relaxation rates $\Gamma_{1,2}$. For example, for frequency range around $30$~Hz one may easy recalculate:
\begin{subequations}
\label{param2}
 \begin{align}
  I_1 &\simeq 18.4\, \text{kW},\quad 
  \Gamma_1 \simeq  0.89\, \text{s}^{-1},\quad \Delta_1\simeq 282\, \text{s}^{-1},\\
  I_2 & \simeq 0.026\, \text{kW},\quad 
  	\Gamma_2 \simeq 4.48\, \text{s}^{-1},\quad \Delta_2 \simeq  -183\, \text{s}^{-1},\nonumber\\
       \Omega_1 &\simeq \pm 2\pi\times 28.3 - i\,0.89\, \text{s}^{-1},\\
       \Omega_2 &\simeq \pm 2\pi\times 30 - i\,1.15\, \text{s}^{-1},\\
       \Omega_3 &\simeq \pm 2\pi\times 34 - i\,3.33\, \text{s}^{-1}.
 \end{align}
\end{subequations} 
This case is more attractive due to modest requirements for powers circulating in arms ($18$~kW instead of $730$~kW). However, estimates above shows tough requirements for bandwidth $\Gamma_1\simeq 0.89\ \text{s}^{-1}$, which is close to limit due to inevitable optical losses. This fact forces us to consider optical losses. 

\begin{figure}[t]
\includegraphics[width = \linewidth]{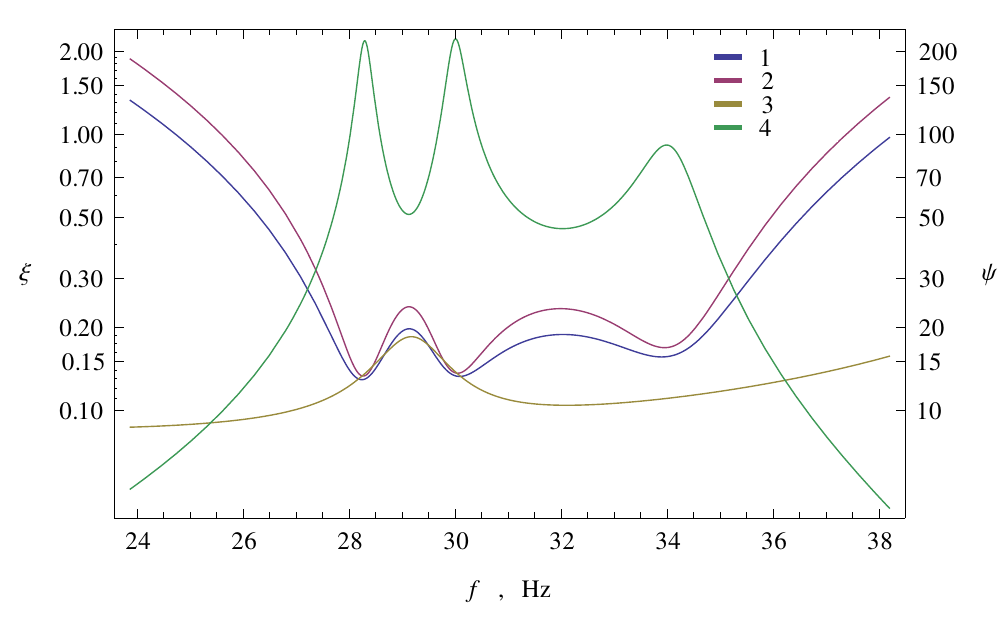}
\caption{Traces 1,2: sensitivities $\xi^{\zeta=0}(f)$ without and with account of optical losses ($\Gamma_A=0$ and $\Gamma_A=\Gamma_1$ correspondingly); 3: sensitivity $\xi_\text{BA}$ (\ref{ShK1}) defined only by back action; 4: dimensionless susceptibility $\psi = \mu(2 \pi f)^2 \Psi$ for parameters~\eqref{param2}.}
\label{fig:plot2}
\end{figure}

\subsubsection{Account of optical losses}

We will take into account only the optical losses in arms' mirrors which may be characterized by round trip loss coefficient $A^2$. We do not take into account losses in beam splitter and SRM because its' influences are negligible small. 

It is known \cite{06a1KhLaVyPRD} that relaxation rates $\tilde \Gamma_{1,2}$ of differential mode of interferometer with optical losses (shown on Fig.~\ref{fig:AdvLIGO}) may be separated into two parts
\begin{equation}
 \tilde\Gamma_{1,2}=\Gamma_{1,2} +\Gamma_A,\quad \Gamma_A= \frac{cA^2}{2L}\,
\end{equation} 
where $\Gamma_1$ describes relaxation through SRM (\ref{Gamma}) (may be called as load relaxation) and $\Gamma_A$ --- relaxation due to optical losses (loss relaxation). 

There is a useful rule~\cite{06a1KhLaVyPRD}: a lossy optical position meter is equivalent to the similar lossless one with gray filter (with power transmission $\Gamma_A/\tilde \Gamma_1$) attached to its signal port. It means that for lossy case, for example, the formula (\ref{starteq}) for output quadrature $\tilde b_{1}^{(1)}$ has to be rewritten in form: 
\begin{align}
\label{tildeb1}
  \tilde b_{1}^{(1)} &= \sqrt{\frac{\Gamma_A}{\Gamma_1+\Gamma_A}}\,e_1^{(1)}+ \\
  &\qquad +\sqrt{\frac{\Gamma_1}{\Gamma_1+\Gamma_A}}\left(a_1^{(1)} +
  	\,U_1\, \mu\Omega^2\Psi\, \mathcal{F} \right)_{\Gamma_{1,2}\to \tilde \Gamma_{1,2}},\nonumber
\end{align}
where quadrature $e_1^{(1)}$ describes additional vacuum fluctuations (with the same spectral density (\ref{Sd})) appearing due to optical losses. Generalization for other quadratures is obvious. Additionally we should replace in all formulas for $\Psi,\ U_i,\ T_i$ relaxation rates $\Gamma_{1,2}$ by $\tilde \Gamma_{1,2}$.

For planned in aLIGO losses (see Table~\ref{table:LIGO}) one may estimate $\Gamma_A\simeq 0.4\ \text{s}^{-1}$. In according with estimate (\ref{param2}) we have to have $\tilde \Gamma_1\simeq 0.89\ \text{s}^{-1}$, i.e. the load and loss relaxations are approximately equal: $\Gamma_1\simeq\Gamma_A$.
 
The corresponding sensitivities plots are presented on Fig.~\ref{fig:plot2}. We see that so dramatically large optical losses practically do not bring any significant degradation to the sensitivity (!). 

So we see that it is possible to circumvent SQL by about $10$ times in wide bandwidth (about the mean frequency) using two optical stable springs.

Note that for simplicity we considered above {\em not optimal} case when only quadrature in one channel is measured. However, this sensitivity only slightly differs from maximal sensitivity when quadratures from both channels with {\em optimal} homodyne angles are measured and then are properly combined --- see details in Appendix~\ref{Details}.  

\subsubsection{Application to Einstein Telescope}

\begin{figure}[t]
	\begin{center}
		\includegraphics[width=\linewidth]{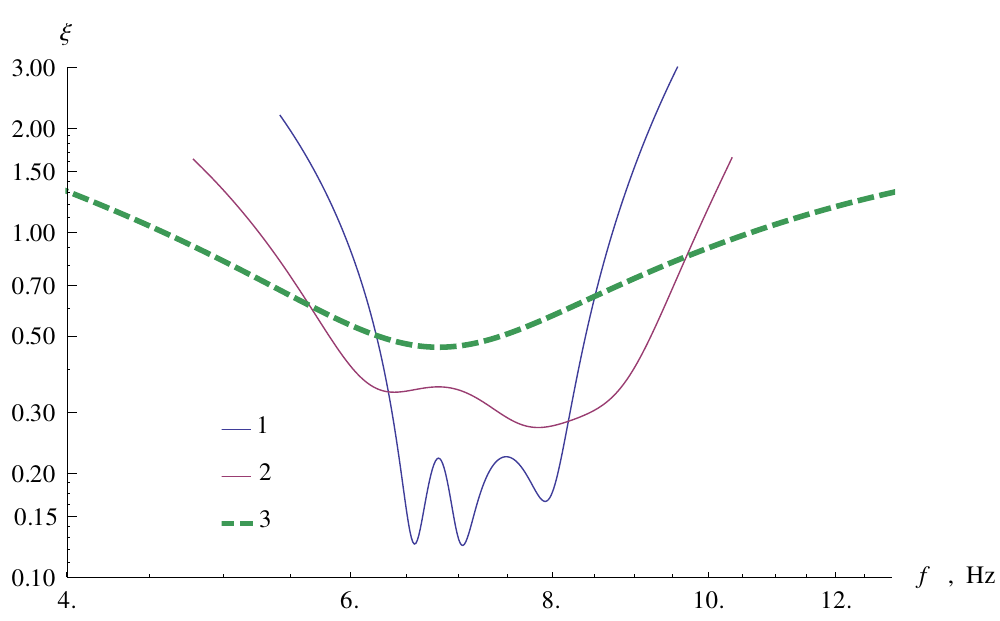}
		\caption{Plots of sensitivity  $\xi$ of ET normalized to SQL versus frequency. Traces~1,2:~different regimes of optical rigidity. Trace~3:~design sensitivity of ET without squeezing, no thermal noise present.}
		\label{fig:et}
	\end{center}
\end{figure}

As the most promising device for gravitational wave detection in future is going to be Einstein Telescope~(ET), it seems useful to try out double optical spring for its configuration. 

We take the parameters for low-frequency setup of ET (see \cite{Hild10}) and recalculate regime from fig.~\ref{fig:plot2} using these parameters and aiming to obtain mean frequency approximately equal to 7 Hz. The result of this recalculation is trace 1 in fig.~\ref{fig:et}. For each of traces 1 and 2 at fig.~\ref{fig:et} we define optimal set of quadratures and filter functions to maximize sensitivity according to algorithm given in appendix~\ref{Details}.

The regime of three close roots however occurs to provide very narrow sensitivity curve as compared with ET design configuration (trace 3 at fig.~\ref{fig:et}). Using procedure described in \cite{11Rakhubovsky} it is quite easy to find some intermediate regimes which demonstrate wider sensitivity curve than trace 1 does, but with peak sensitivity better then one designed for ET. We present one such regime plotted by means of trace 2 at fig.~\ref{fig:et}. These regimes mainly differ by detunings and relaxation rates of main and auxiliary pumps. Required values of power in main pump are approximately equal for traces 1 and 2. Also these values are smaller than power in design configuration (18 kW). However power of auxiliary pump changes significantly (still remaining small as compared to power of main pump) while sensitivity curve becomes wider. For details see Appendix~\ref{sec:tuning}.

\subsection{Speculations on quantum behavior of optomechanical degrees of freedom}

Let remind that oscillator in thermal bath at temperature $T$  will demonstrate quantum behavior  \cite{1968_SQL,1992_quant_meas,1996_QND_toys_tools} if the following inequality is fulfilled
\begin{align}
\label{Quantum}
 \frac{\varkappa T}{\hslash \Omega \,Q} <1, 
\end{align}
where $\varkappa$ is Boltzmann constant, $\Omega$ and $Q$  are frequency and quality factor of oscillator.

The plots of susceptibility on Figs.\ref{fig:plot1}, \ref{fig:plot2} show three peaks corresponding to three optomechanical degrees of freedom. It is important that they are {\em stable} ones. Generally speaking  it is not clear how to describe correctly these degrees of freedom similar to normal coordinates in case of coupled oscillators. However, for estimation one can consider the mechanical degree of freedom and calculate mean kinetic and potential energy $E_i$ stored in each peak:
 \begin{align}
  \int_{\Omega_i-3\Delta \Omega_i}^{\Omega_i+3\Delta \Omega_i} \frac{\big[\mu\Omega^2+\Re(K_1+K_2)\big]|z(\Omega)|^2}{2}\, \frac{d\Omega}{2\pi}
 \end{align}
where $\Omega_i$ and $\Delta\Omega_i$ are mean frequency and bandwidth of each peak (real and imaginary parts of roots (\ref{roots1},\ref{param2})). Let assume that only fluctuational light pressure forces originate fluctuations of displacement $z$ (no thermal noise). The numerical calculations gives the following values:
\begin{align}
 E_1 &\simeq 1.28\, \hslash \Omega_1, \quad E_2\simeq 1.76\, \hslash \Omega_2,\quad
 	E_3\simeq 1.1\, \hslash \Omega_3
\end{align}
We see that mean energy in each degree of freedom is close to one quanta, hence, they should demonstrate quantum behavior. It becomes obvious if we apply criteria (\ref{Quantum}) for each peak:
\begin{align}
 \frac{E_1}{\hslash \Omega_1\, Q_1}&\simeq 0.012\ll 1,\\
 \frac{E_2}{\hslash \Omega_2\, Q_2}&\simeq 0.024\ll 1,\\
 \frac{E_3}{\hslash \Omega_3\, Q_3}&\simeq 0.034 \ll 1, \quad Q_i\equiv \frac{\Omega_i}{2\Delta \Omega_i}
\end{align}

It provides incredible possibility to observe quantum behavior of really macroscopic object with mass $\mu\simeq 10$~kg~(!)

\section{Conclusion}

We have shown that usage of {\em stable} double optical spring regime in laser gravitational detectors may provide sensitivity gain if relaxation rates of optical modes are much smaller than detunings. However, the decrease of  relaxation rates is restricted by optical losses in mirror. We have shown that even for case when relaxation through signal recycling mirror (load relaxation) and  relaxation via optical losses (loss relaxation) are equal to each other the degradation of sensitivity is relatively small. 

The presented example of close resonances regime via double pumped optical spring promises the possibility to circumvent Standard Quantum Limit by about ten times in the frequency range about half of mean frequency. Experimenter may further vary parameters set (pump powers, detunings) to control susceptibility and, hence, sensitivity curve. Of course, this gain of sensitivity will take place if the level of thermal and technical noises is low enough. The thermal noise in mirror's coating makes the main contribution in thermal noise budget. However, one may hope for progress in manufacture of the interferometric coating. Note that the planned level of thermal noise in laser gravitational detectors of third generation (Einstein Telescope ET-D Low Frequency Interferometer) is planned to be about ten times less than Standard Quantum Limit \cite{et_hild2011} in frequency range about $10$ Hz. 

If thermal noise is small enough the stable double optical spring provides unique possibility to observe quantum behavior of macroscopic object with effective mass about 10 kg, because fluctuational energy (created by back action force) stored in each peak of sensitivity is about one quanta $\hslash \Omega_i$. 

\acknowledgments
The authors are grateful for fruitful discussions to Stefan Hild. We  also thank Yanbei Chen, Farid Khalili and Stefan Danilishin. Authors are supported  by the Russian Foundation for Basic Research Grant No. 08-02-00580-a and NSF grant  PHY-0967049. 

\appendix

\section{Initial formulas}\label{notations}
In this Appendix we derive formulas (\ref{starteq}, \ref{UT}). The electric fields $E_{1,2}$ in propagating wave of each pump and corresponding the mean intensities $ J_{1,2}$ of light beam can be written as follows \cite{02a1KiLeMaThVyPRD}: 
\begin{align*}
E_{1,2} & \simeq  \sqrt{\frac{2\pi \, \hslash \omega_{1,2}}{Sc} }\, e^{-i\omega_{1,2} t}\times\\
        &\quad \times \left( \tilde A_{1,2} +\int_{-\infty}^\infty
         a_{1,2}(\Omega)\, e^{-i\Omega t} \frac{d\Omega}{2\pi}\right)
         +  \big\{\text{h.c.} \big\},\nonumber\\
J_{1,2} & =  \hslash \omega_{1,2}|\tilde A_{1,2}|^2,\quad
        \big[a_{1,2}(\Omega),\,a_{1,2}^+(\Omega')\big]=2\pi\,\delta(\Omega -\Omega'),
\end{align*}
where $S$ is the cross section of the light beam, $c$ is the velocity of light, $a$ and $a^+$ are annihilation and creation operators.   

One can obtain the formula in frequency domain for the output fields $b_{1,2}$ in dark port as a function of the input fluctuational field $a_{1,2}$ and mirror positions  \cite{08ChenPRD}:
\begin{align}
\label{b12}
 b_{1,2}&\,G_{1,2}^* = \frac{\Gamma_{1,2}+i(\Omega +\Delta_{1,2}) }{
	\Gamma_{1,2} -i(\Omega+\Delta_{1,2})}\,
	  a_{1,2}\,G_{1,2} +\\
	&\quad +\sqrt\frac{2\omega_{1,2}\Gamma_{1,2}I_{1,2}}{\hslash cL}\, \frac{ i\, z}{
     \Gamma_{1,2} -i(\Omega+\Delta_{1,2}) },\nonumber\\
z = & \big(x_N - y_N\big) - \big(x_E- y_E\big),\quad G_{1,2} \equiv 
	\sqrt \frac{e^{2i\phi_{1,2}}+\rho}{1+\rho e^{2i\phi_{1,2}}}\,.\nonumber
\end{align}
where  $x_E,\ y_E,\ x_N, \ y_N$ --- are the displacements of FP mirrors (see notations on
Fig.\ref{fig:AdvLIGO}),  $I_{1,2}$ are mean optical power circulating in each arm from pumps $1$ or $2$. 

As we assume that input fields are in vacuum state (no squeezing), it is convenient to denote
\begin{subequations}
 \label{bara}
\begin{align}
\bar b_{1,2} &=b_{1,2} \,G_{1,2}^*, \\
	\bar a_{1,2}&=\frac{\Gamma_{1,2}+i(\Omega +\Delta_{1,2}) }{
	\Gamma_{1,2} -i(\Omega+\Delta_{1,2})}\,a_{1,2}\,G_{1,2},
\end{align}
\end{subequations}

As usual we introduce input  $a_{1,2}^{(1,2)}$ and output $b_{1,2}^{(1,2)}$ quadrature amplitudes
\begin{subequations}
\label{newQ}
\begin{align}
a_{1,2}^{(1)} &\equiv \big[\bar a_{1,2}(\Omega)+\bar a_{1,2}^+(-\Omega)\big]/\sqrt 2,\\
a_{1,2}^{(2)} &\equiv \big[\bar a_{1,2}(\Omega) - \bar a_{1,2}^+(-\Omega)\big]/i\sqrt 2,\\
b_{1,2}^{(1)} &\equiv \big[\bar b_{1,2}(\Omega) + \bar b_{1,2}^+(-\Omega)\big]/\sqrt 2,\\
b_{1,2}^{(2)} &\equiv \big[\bar b_{1,2}(\Omega)-\bar b_{1,2}^+(-\Omega)\big]/i\sqrt 2.
\end{align} 
\end{subequations}

Now we have to account  evolution of differential coordinate $z$ through susceptibility $\Psi$ in frequency domain:
\begin{align}
\label{z}
z =& \Psi(\Omega)\big(f_1+f_2 +F_s \big),\quad F_s= \mu\Omega^2 L\, h   .
\end{align}
Here $F_s$ is the equivalent signal force. Back action forces $f_{1,2}$  produced by fluctuations of light pressure are equal to: 
\begin{align}
\label{f12}
 f_{1,2} &=\sqrt\frac{2\hslash \omega_{1,2}\Gamma_{1,2}I_{1,2}}{c L}\times\\
 	&\qquad \times\left(
 		\frac{\bar a_{1,2}(\Omega)}{\Gamma_{1,2}+i\Delta_{1,2}+i\Omega}+
 		\frac{\bar a_{1,2}^+(-\Omega)}{\Gamma_{1,2}-i\Delta_{1,2}+i\Omega}\right).\nonumber
\end{align}
Substituting (\ref{b12}, \ref{bara}, \ref{z}, \ref{f12}) into definition (\ref{newQ}) of quadrature amplitudes one may obtain the formulas (\ref{starteq}, \ref{UT}).

\section{Accurate sensitivity calculation}\label{Details}

The sensitivity of GW-antenna depends on homodyne angles and the filter functions that are used during procession of quadratures and homodyne currents. The optimal set of filter functions can be found as the eigen vector of the proper matrix \cite{68Korn,08ChenPRD}. Below we provide the algorithm to obtain this set in general case of multiple pumps and calculate optimal sensitivity for the case of two pumps.

Consider that measurement of output quadratures gives several homodyne currents $j_i$ (defined by formulae~\eqref{b12psi} in case of two pumps), each of which consists of signal part $j_i^{(S)}$ and noise part $j_i^{(N)}$. The total output is built as the weighted sum of these currents with filter functions $Y_i$: 
\begin{equation}
	\notag
	J = \sum_i Y_i j_i.
\end{equation}
Spectral density of signal $S_s$ and noise $S_n$ parts of output signal can be presented as the quadratic forms over the vector of filter functions $\vec Y \equiv \{ Y_1; Y_2; \dots \}$. 
\begin{align}
	\notag
	S_n & = (\vec Y; \hat N \vec Y) \equiv \sum \limits_{ij} Y^*_i N_{ij} Y_j;
	\\
	\notag
	S_s & = (\vec Y; \hat S \vec Y) \equiv \sum \limits_{ij} Y^*_i S_{ij} Y_j;
\end{align}

\begin{figure}[t]
	\includegraphics[width = \linewidth]{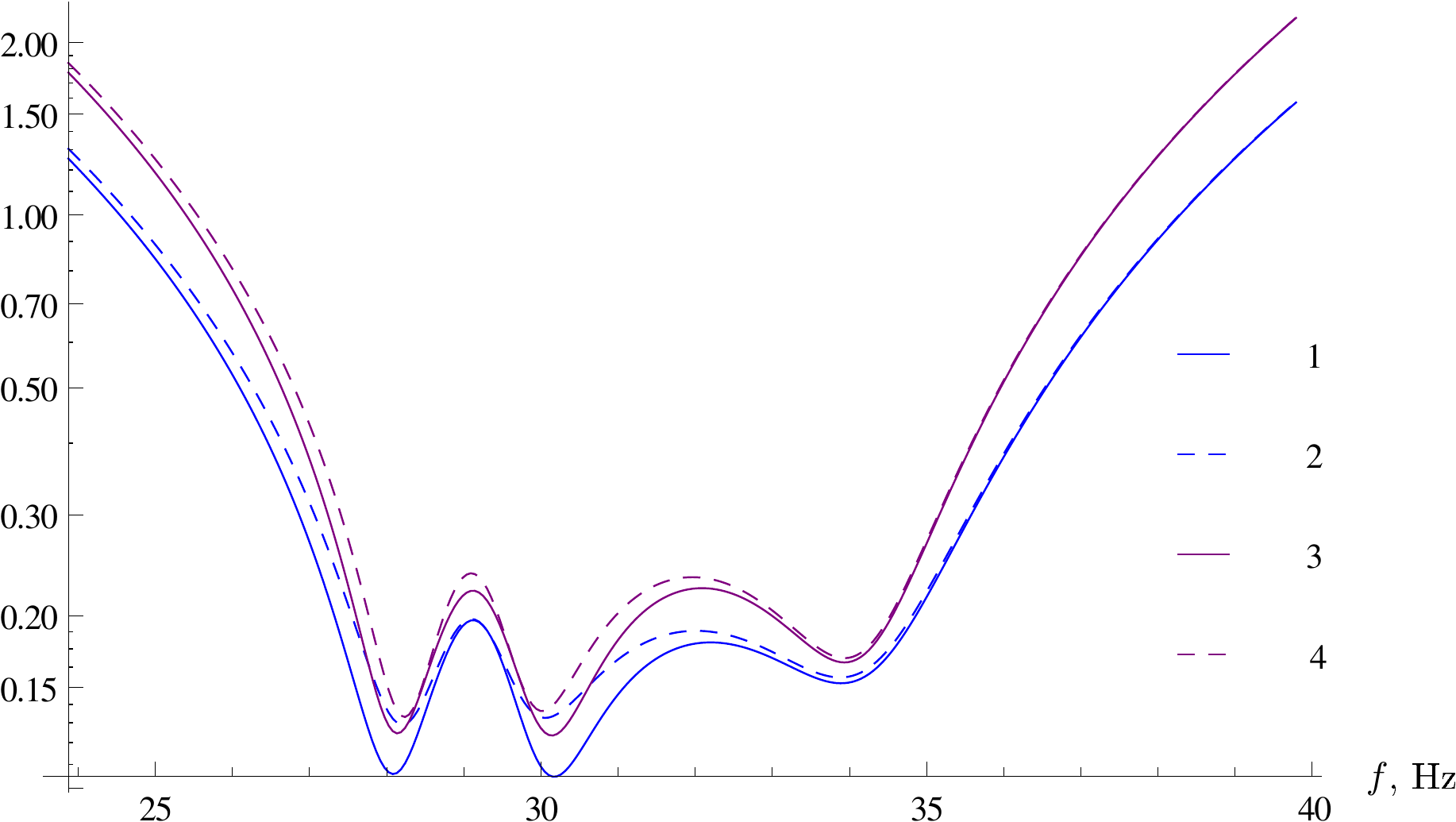}
	\caption{Plots of sensitivity versus frequency. 1: optimized sensitivity $\xi_i(f)$, losses are absent; 2: sensitivity $\xi^{\zeta=0}(f)$ in case of measurement of only the amplitude quadrature of the strong pump, losses are absent. 3 and 4: the same with losses $\Gamma_A = \Gamma_1$. }
	\label{fig:sensloss}
\end{figure}

The elements of matrices $\hat S$ and $\hat N$ of the corresponding quadratic forms are defined by expressions
\begin{equation}
	S_{ij} (\Omega,\vec \zeta) = \overline{ j_i^{(S)*} j_j^{(S)} }; \quad 	N_{ij} (\Omega, \vec \zeta) = \overline{ j_i^{(N)*} j_j^{(N)} },
	\label{eq:SNmatr-def}
\end{equation}
where the line over means the operation of calculation of the cross spectral density. These elements depend on frequency $\Omega$ and homodyne angles used in quadratures measurements. 

Inverse sensitivity of antenna is given as the ratio 
\begin{equation}
	\xi^{-1} = \frac{ 1}{h^2_\text{SQL}} \frac{ S_s }{ S_n } = \xi^{-1} ( \Omega; \vec \zeta; \vec Y ). 
	\notag
\end{equation}
It depends on the same quantities and also on the set of filter functions. The optimal set of the latter should maximize $\xi^{-1}$ thus minimizing the sensitivity. 

The eigen vector of the matrix $ \hat P \equiv \hat N^{-1} \cdot \hat S$ that corresponds to it's largest eigen value provides desired set of filter functions \cite{68Korn,08ChenPRD}. Corresponding sensitivity is given by inverse to that eigen value. 

The optimal sensitivity calculated following this method depends on frequency and homodyne angles. Optimization (minimization) over the latter gives optimal sensitivity $\xi_o (\Omega)$ for the case of frequency dependent homodyne angles. Due to experimental difficulties arising of realisation of such regime it is worth to make estimation for optimal set of homodyne angles.

As a criterion we use the condition of minimization of integral sensitivity $\mathcal S$
\begin{equation}
	\notag
	\mathcal S (\vec \zeta) = \int \limits_{\Omega_1}^{\Omega_2} \xi (\Omega,\vec \zeta) d\Omega,
\end{equation}
where frequencies $\Omega_{1;2}$ are boundaries of the range in which the sensitivity $\xi^{\zeta=0} (\Omega)$ is under the SQL level. This optimization gives the best sensitivity $\xi_{i}$ acheivable with frequency independent homodyne angles.

In the case of double pump all the matrices and vectors are two-dimensional so all the eigen vectors and eigen values can be calculated analytically. Further optimization over the homodyne angles is done numerically. Results are presented in the fig. \ref{fig:sensloss}. One can see that that sensitivity does not differ dramatically as compared with the approximation presented in Sec.~\ref{dpumpOut} (taking into account only output from one pump).

\section{Tuning details for fig.~\ref{fig:et}}\label{sec:tuning}

Parameters needed to achieve regimes with sensitivities represented by first two traces at fig.~\ref{fig:et} are listed in Table~\ref{tab:tuning} below. 

\begin{table}[h]
	\caption{Parameters for plots at fig.~\ref{fig:et}}\label{tab:tuning}
\begin{tabular}{r|c|c}
	& Trace 1 & Trace 2 
	\\
	\hline
	$I_1$, kW & 4.5 & 5.4 
	\\
	$I_2$, kW & 0.06 & 0.12 
	\\
	$\Delta_1$, s$^{-1}$ & 65.7 & 70.5 
	\\
	$\Delta_2$, s$^{-1}$ & -42.8 & -43.8 
	\\
	$\Gamma_1$, s$^{-1}$ & 0.21 & 1.05 
	\\
	$\Gamma_2$, s$^{-1}$ & 1.04 & 6.70 
\end{tabular}
\end{table}


\end{document}